\documentclass[aps, showpacs, showkeys,nofootinbib,floatfix]{revtex4}

\usepackage{amssymb}
\usepackage{amsmath}
\usepackage{graphicx}
\usepackage{hyperref}

\begin{document}

\title{The Real Scalar Field Equation for Nariai Black Hole in the 5D Schwarzschild-de Sitter Black String Space}
\author{Molin Liu}
\email{mlliudl@student.dlut.edu.cn}
\author{Hongya Liu}
\email{hyliu@dlut.edu.cn}
\author{Chunxiao Wang}
\author{Yongli Ping}

\affiliation{School of Physics and Optoelectronic Technology,
Dalian University of Technology, Dalian, 116024, P. R. China}

\begin{abstract}
The Nariai black hole, whose two horizons are lying close to each
other, is an extreme and important case in the research of black
hole. In this paper we study the evolution of a massless scalar
field scattered around in 5D Schwarzschild-de Sitter black string
space. Using the method shown by Brevik and Simonsen (2001) we
solve the scalar field equation as a boundary value problem, where
real boundary condition is employed. Then with convenient
replacement of the 5D continuous potential by square barrier, the
reflection and transmission coefficients ($R,\ T$) are obtained.
At last, we also compare the coefficients with usual 4D
counterpart.
\end{abstract}



\maketitle

\section{Introduction}
Black hole radiation, which was proved originally by Stephen
Hawking with the method of quantum field on gravitational
collapsing \cite{ref:Hawking1} \cite{ref:Hawking2}, indicates that
black holes are not perfect black but radiate thermally and
eventually explode. Since then, many people have used various
methods and techniques to research black hole through the
particles radiating from it, such as the simple Klein-Gordon
particles and Dirac particles (for some early works, see Damour
and Ruffini \cite{ref:Damour} and Chandrasekhar
\cite{ref:Chandrasekhar} respectively). Here, scalar particles are
only considered. Recently, Higuchi et al. \cite{ref:Higuchi} and
Grispino et al. \cite{ref:Grispino} gave the scalar field solution
outside a Schwarzschild black hole; Brady et al. \cite{ref:Brady},
Brevik et al. \cite{ref:Brevik} and Tian et al. \cite{ref:Tian}
studied the Schwarzschild-de Sitter (SdS) case; Guo et al.
\cite{ref:Guo} made further studies in the
Reissner-Nordstr\"{o}m-de Sitter one. As for recent studies of
searching evaporating black holes, one can refer to the works
\cite{ref:Barrau1} \cite{ref:Barrau2} \cite{ref:Carr}.

The idea that the world may have more than four dimensions is due
to Kaluza \cite{ref:Kaluza} and Klein \cite{ref:Klein}, who
realized that a 5D manifold could be used to unify general
relativity with Maxwell's theory of electromagnetism. After that,
many people focus on the robust higher dimensional space. Here, we
consider the Space-Time-Matter (STM) theory presented by Wesson
and co-workers
 \cite{ref:Wesson} \cite{ref:Overduin}. This theory is distinguished from the
classical Kaluza-Klein theory for a non-compact fifth dimension,
the 4D source is induced from an empty 5D manifold. Because of
this, the STM theory is also called induced matter theory and the
effective 4D matter is called induced matter. That is, in STM
theory, 5D manifold is Ricci-flat while 4D hypersurface is curved
by the 4D induced matter. Mathematically, this approach is
supported by Campbell's theorem which states that any analytical
solution of N-dimensional Einstein equations with a source can be
locally embedded in an (N+1)-dimensional Ricci-flat manifold
\cite{ref:Campbell}. In the framework of STM, people studied many
works such as Quantum Dirac Equation \cite{ref:Macias}, Perihelion
Problem \cite{ref:Lim}, Kaluza-Klein Solitons \cite{ref:Billyard},
Black Hole \cite{ref:Liu222} \cite{ref:Liu00}, Solar System Tests
\cite{ref:Liu333} and so on.

In order to avoid interactions beyond any acceptable
phenomenological limits, people assume standard model fields (such
as fermions, gauge bosons, Higgs fields) are confined on a ($3+1$)
dimensional hypersurface (3-brane) without accessing along the
transverse dimensions. The branes are embedded in the higher
dimensional spacetime (bulk), in which only gravitons and scalar
particles without charges could propagate under standard model
gauge group. There are also many works (for a review with large
extra dimensions see \cite{ref:Kantis}) focusing on Hawking
radiation such as \cite{ref:Kanti3} \cite{ref:Kanti2}
\cite{ref:Duffy} \cite{ref:Kanti4} \cite{ref:Kanti5}.

Cosmological constant $\Lambda$, which is a parameter with
dimension $L^{-2}$ ($L$ is length), is one of focuses in
Gravitation Theory. The acceleration of the universe is explained
by the required repulsive force produced by a non-zero and
positive cosmological constant $\Lambda$. Current SnIa observation
data shows that the cosmological constant has a value of
$\Lambda_{0} \sim 10^{-52} m^{-2}$ \cite{ref:Peebles}
\cite{ref:Kagramanova} \cite{ref:Axenides}. Its robust non-zero
magnitude engages the researching interest in the space contained
cosmological constant. Especially, the black hole contained
effective cosmological constant is studied widely either in higher
dimensions background \cite{ref:Kantis} \cite{ref:Kanti3}
\cite{ref:Liu00} or in usual 4D case \cite{ref:Brady}
\cite{ref:Brevik} \cite{ref:Tian} \cite{ref:Guo}. Sometimes for
the sake of study, $\Lambda$ is considered as a free parameter
like in the works \cite{ref:Brevik} \cite{ref:Tian}
\cite{ref:Guo}. In SdS space, the interval between black hole
horizon $r_{e}$ and cosmological horizon $r_{c}$ becomes smaller
with increase value of $\Lambda$. If cosmological constant
$\Lambda$ reaches its maximum, Nariai black hole will be arisen.
In this paper, we study how a massless scalar field evolves in
this extreme case.

This paper is organized as follows: in section II, the 5D SdS
black string space, the time-dependent radial equation about
$R_{\omega}(r,t)$ and the fifth dimensional equation about $L(y)$
are restated. In section III, by a tortoise coordinate
transformation, the radial equation becomes a Schr\"{o}dinger
wavelike one. According to the boundary condition and the tangent
approximation, a full numerical solution is presented. In section
VI, using the replacement of real potential barriers around black
hole by square barriers, the reflection and transmission
coefficients are obtained. Section V is a conclusion.

We adopt the signature (+, -, -, -, -), put $\hbar$, $c$, and $G$
equal to unity. Lowercase Greek indices $\mu$, $\nu$, $\ldots$
will be taken to run over 0, 1, 2, 3 as usual, while capital
indices A, B, C $\ldots$ run over all five coordinates
(0,1,2,3,4).
\section{The Massless Scalar Field in 5D Schwarzschild-de Sitter Black String Space}
Within the framework of STM theory, an exact 5D solution presented
by Mashhoon, Wesson and Liu \cite{ref:Wesson} \cite{ref:Mashhoon}
\cite{ref:Liu} describes a 5D black hole. The line element takes
the form
\begin{equation}
dS^{2}=\frac{\Lambda
\xi^2}{3}\left[f(r)dt^{2}-\frac{1}{f(r)}dr^{2}-r^{2}\left(d\theta^2+\sin^2\theta d\phi^2\right)\right]-d\xi^{2}. \label{eq:5dmetric}%
\end{equation}
In our case
\begin{equation}
f(r)=1-\frac{2M}{r}-\frac{\Lambda}{3}r^2,\label{f-function}
\end{equation}
where $\xi$ is the open non-compact extra dimension coordinate,
$\Lambda$ is the induced cosmological constant and $M$ is the
central mass. The part of this metric inside the square bracket is
exactly the same line-element as the 4D SdS solution, which is
bounded by two horizons
--- an inner horizon (black hole horizon) and an outer horizon (one may call it cosmological horizon). This metric
(\ref{eq:5dmetric}) satisfies the 5D vacuum equation $R_{AB}=0$,
therefore, there is no cosmological constant when viewed from 5D.
However when viewed from 4D, there is an effective cosmological
constant $\Lambda$. So one can actually treat this $\Lambda$ as a
parameter which comes from the fifth dimension. This solution has
been studied in many works \cite{ref:Mashhoon11} focusing mainly
on the induced constant $\Lambda$, the extra force and so on.

We redefine the fifth dimension in this model,
\begin{equation}
\xi=\sqrt{\frac{3}{\Lambda}}e^{\sqrt{\frac{\Lambda}{3}}y}.\label{replacement}
\end{equation}
Then we use (\ref{eq:5dmetric}) $\sim$ (\ref{replacement}) to
build up a RS type brane model in which one brane is at $y=0$, and
the other brane is at $y=y_{1}$. Hence the fifth dimension becomes
finite. It could be very small as RS I brane model
\cite{ref:Randall2} or very large as RS II model
\cite{ref:Randall1}. The relation between STM theory and brane
world theories, and the embedding of 5D solutions to brane models
are studies in \cite{Ponce} \cite{Seahra} \cite{Liu_plb}
\cite{Ping}. For the present brane model, when viewed from a
($\xi$ or $y$ = $constant$) hypersurface, the 4D line-element
represents exactly the SdS black hole. However, when viewed from
5D, the horizon does not form a 4D sphere --- it looks like a
black string lying along the fifth dimension. Usually, people call
the solution to the 5D equation $^{(5)}G_{AB}$ = $\Lambda _{5}$$
^{(5)}g_{AB}$ ($\Lambda_{5}$ is the 5D cosmological constant) as
the 5D SdS solution. Therefore, to distinguish with it, we call
the solution (\ref{eq:5dmetric}) a black string, or more
precisely, a 5D Ricci-flat SdS solution.

After redefining the fifth dimension, the metric
(\ref{eq:5dmetric}) can be rewritten as
\begin{equation}
dS^{2}=e^{2\sqrt{\frac{\Lambda}{3}}y}\left[f(r)dt^{2}-\frac{1}{f(r)}dr^{2}-r^{2}\left(d\theta^2+\sin^2\theta d\phi^2\right)-dy^{2}\right],\label{eq:5dmetric-y}%
\end{equation}
where $y$ is the new fifth dimension. Expression
(\ref{f-function}) can be recomposed as follows
\begin{equation}
f(r)=\frac{\Lambda}{3r}(r-r_{e})(r_{c}-r)(r-r_{o}). \label{re-f function}%
\end{equation}

The singularity of the metric (\ref{eq:5dmetric-y}) is determined
by $f(r)=0$. Here we only consider the real solutions. The
solutions to this equation are black hole event horizon $r_{e}$,
cosmological horizon $r_{c}$ and a negative solution
$r_{o}=-(r_{e}+r_{c})$. The last one has no physical significance,
and $r_{c}$ and $r_{e}$ are given as

\begin{equation}
\left\{
\begin{array}{c}
r_{c} = \frac{2}{\sqrt{\Lambda}}\cos\eta ,\\
r_{e} = \frac{2}{\sqrt{\Lambda}}\cos(120^\circ-\eta),\\
\end{array}
\right.\label{re-rc}
\end{equation}
where $\eta=\frac{1}{3}\arccos(-3M\sqrt{\Lambda})$ with $30^\circ
\leq\eta\leq 60^\circ$. The real physical solutions are accepted
only if
 $\Lambda$ satisfy $\Lambda M^2\leq\frac{1}{9}$ \cite{ref:Liu}.

Then we consider a massless scalar field $\Phi$ in the 5D black
string spacetime, obeying the Klein-Gordon equation
\begin{equation}
\square\Phi=0,\label{Klein-Gorden equation}%
\end{equation}
where $ \square=\frac{1}{\sqrt{g}}\frac{\partial}{\partial
x^{A}}\left(\sqrt{g}g^{AB}\frac{\partial}{\partial{x^{B}}}\right)\label{Dlb}
$  is the 5D d'Alembertian operator. We suppose that the separable
solutions to Eq. (\ref{Klein-Gorden equation}) are in the form
\begin{equation}
\Phi=\frac{1}{\sqrt{4\pi\omega}}\frac{1}{r}R_{\omega}(r,t)L(y)Y_{lm}(\theta,\phi),\label{wave
function}
\end{equation}
where $R_{\omega}(r,t)$ is the radial time-dependent function,
$Y_{lm}(\theta,\phi)$ is the usual spherical harmonic function,
and $L(y)$ is the function about the fifth dimension. The
differential equations about $y$ and (t, r) are
\begin{eqnarray}
 &&\frac{d^2L(y)}{dy^2}+\Lambda\sqrt{\frac{\Lambda}{3}}\frac{d L(y)}{dy}+\Omega
 L(y)=0,\label{5-th-equation} \\
 && -\frac{1}{f(r)} r^2\frac{\partial^2}{\partial
 t^2}\left(\frac{R_{\omega}}{r}\right)+\frac{\partial}{\partial r}\left(r^2
 f(r)\frac{\partial}{\partial{r}}\left(\frac{R_{\omega}}{r}\right)\right)-\left[\Omega r^2+l(l+1)\right]\frac{R_{\omega}}{r}=0.\label{radius-t-equation}
  \end{eqnarray}
 Eq. (\ref{radius-t-equation}) is a time-dependent radial differential equation. Eq. (\ref{5-th-equation}) is a differential equation about $y$, where
 $\Omega$ is a constant which is adopted to separate variables $(t,
 r,\theta,\phi, y )$.

\section{The Nariai black hole and its boundary value problem}
Nariai black hole \cite{ref:Nariai} \cite{ref:Nojiri} occurs when
the cosmological horizon is very close to the black hole horizon
$r_{c}\longrightarrow r_{e}$. It is an extreme and important kind
of SdS black holes. The cosmological constant in this limit is
given by
\begin{equation}\label{cosmolo-limit}
    \Lambda M^{2}=\frac{1}{9}.
\end{equation}
Substituting Eq. (\ref{cosmolo-limit}) into Eq. (\ref{re-rc}), we
can get $\eta=60^\circ$ and the horizons $r_{h}=r_{e}=r_{c}=3M$.
As an illustration of the accuracy , we mention that the choice
$\Lambda M^2=0.11$ \cite{ref:Brevik} \cite{ref:Tian} leads to
$r_{e}=2.8391M$ and $r_{c}=3.1878M$. In order to simplify
numerical calculation, we will put $M = 1$ in this paper.
\subsection{The Fifth Dimensional Function L(y)}

In our previous paper \cite{ref:Liu00}, we have introduced a
massless scalar field to stabilizing this black string brane
model. Considering a single mode of the scalar field, the wave
function for this mode may reach its maximum value but keep smooth
and finite at the brane. Hence, a steady standing wave is
constructed. A suitable superposition of some of the quantized and
continuous components of $L(y)$ may provide a wave function which
is very large at $y=0$ and drops rapidly for $y \neq 0$.
Naturally, a practical 3-brane is formed at the $y=0$
hypersurface. According to this $''\text{standing wave}''$
condition in the bulk, the spectrum of $\Omega$ is broken into two
parts. One is the continuous spectra below $\frac{3}{4}\Lambda$
and the other is the discrete spectra above $\frac{3}{4}\Lambda$.
The quantum parameter $\Omega_{n}$ is
\begin{equation}\label{quan-Omega}
    \Omega_{n}=\frac{n^2\pi^2}{y_{1}^2}+\frac{3}{4}\Lambda,
\end{equation}
where n=1, 2, 3 \ldots and $y_{1}$ is the thickness of the bulk.
So the solutions to Eq. (\ref{5-th-equation}) are
\begin{equation}
L(y)=\left\{
\begin{array}{c}
C e^{-\frac{\sqrt{3\Lambda}}{2}y} \cos \left(n \pi \frac{y}{y_{1}} \right),{\ \ \ } n=1,2,3\cdots,{\ }\text{\ for }\Omega >\frac{3\Lambda}{4},\\
\left(C_{1}+C_{2}y\right)e^{-\frac{\sqrt{3\Lambda}}{2}y},{ \ \ \ \ \ \ \ \ \ \ \ \ \ \ }{\ \ \ \ \ \ \ \ \  \ \ \ \ }\text{\ for }\Omega=\frac{3\Lambda}{4},\\
C_{3}e^{\frac{-\sqrt{3\Lambda}+\sqrt{3\Lambda-4\Omega}}{2}y}+C_{4}e^{\frac{-\sqrt{3\Lambda}-\sqrt{3\Lambda-4\Omega}}{2}y},\text{\ \ \ \ for }\Omega<\frac{3\Lambda}{4},\\

\end{array}
\right.\label{5-d-function}
\end{equation}
where $\Lambda=0.11$ is the cosmological constant, and $(C,
y_{0})$, $(C_{1}, C_{2})$, $(C_{3}, C_{4})$ are the three pairs of
integration constants.

\begin{figure}
  \includegraphics[width=3.5 in]{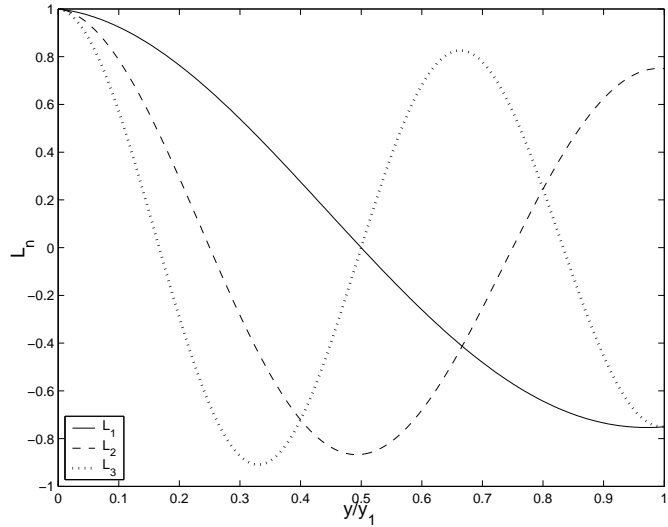}\\
  \caption{The first three eigenfunctions of $L_{n}$(y) in Nariai case: $L_{1}$(solid), $L_{2}$(dashed), and $L_{3}$(dotted) with $y_{1}\rightarrow 1$, C=1 and $\Lambda=0.11.$}\label{quan-states}
\end{figure}

As an illuminating example, Fig. \ref{quan-states} depicts the
quantized sates of $L(y)$. It illustrate that the eigenfunctions
get the maximum on the brane $y = 0$ and get an extremum on the
other brane $y=y_{1}$. For an exponential factor
$e^{-\frac{\sqrt{3\Lambda}}{2}y}$ in the first solution of Eqs.
(\ref{5-d-function}), the wave function $L_{n}$ decays along the
fifth dimension. Comparing with general result \cite{ref:Liu00},
extreme cosmological constant $\Lambda=0.11$ gives a more fiercely
decay solutions. As the fifth dimension becomes bigger, the
probability $|L(y)|^2$ deflects from the original value
($|L(y)|^2_{y = 0} = 1$) more and more larger. It gets an extremum
instead of the maximum value on the other brane. In this way, two
branes can be stabilized by scalar field.

\subsection{The Schr\"{o}dinger wavelike equation}

A more important aspect of scalar field is radial direction. In
Eq. (\ref{radius-t-equation}) time variable can be eliminated by
the Fourier component $e^{-i \omega t}$ via
 \begin{equation}
R_{\omega}(r,t)\rightarrow\Psi_{\omega l n} (r) e^{-i \omega t},
 \end{equation}
 where the subscript $n$ presents a new wave function unlike the usual 4D case $\psi_{\omega l}$. So Eq. (\ref{radius-t-equation}) can be rewritten as%
 \begin{equation}
 \left[-f(r)\frac{d}{dr}(f(r)\frac{d}{dr})+V(r)\right]\Psi_{\omega
 l n}(r)=\omega^2\Psi_{\omega l n} (r),\label{radius equ. about r}
 \end{equation}
where the potential function is given by%
\begin{equation}
V(r)=f(r)\left[\frac{1}{r}\frac{df(r)}{dr}+\frac{l(l+1)}{r^2}+\Omega\right].\label{potential
of r}
\end{equation}

\begin{figure}[tbh]
\centering
\includegraphics[width=3.5in]{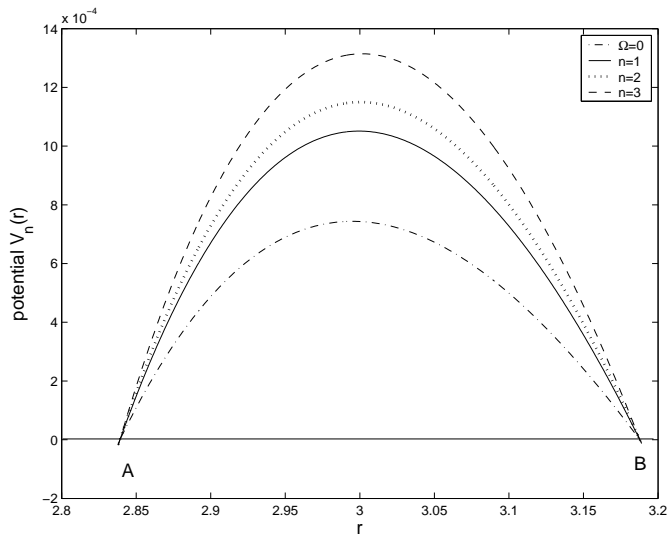}
\caption{The potentials of Nariai black hole with n=1 (solid), n=2
(dotted), and n=3 (dashed). Here M=1, $\Lambda=0.11$, $l=1$,
$y_{1}=10^{3/2}$ (a very large 5th dimension). The usual 4D
potential ($\Omega=0$) is also plotted with dash-dot line for
comparison. The black hole horizon locates at the point A
$r_{e}\sim 2.8391$ and the cosmological horizon locates at the
point B. The potential tends to zero exponentially quickly as
$x\rightarrow\pm\infty.$}\label{fig:quan-potential-r}
\end{figure}

Now we introduce the tortoise coordinate%
\begin{equation}
x=\frac{1}{2M}\int\frac{dr}{f(r)}.\label{tortoise }
\end{equation}
The tortoise coordinate can be expressed by the gravitation
surface as follows
\begin{equation}
x=\frac{1}{2M}\left[\frac{1}{2K_{e}}\ln\left(1-\frac{r}{r_{e}}\right)-\frac{1}{2K_{c}}\ln\left(1-\frac{r}{r_{c}}\right)+\frac{1}{2k_{o}}\ln\left(1-\frac{r}{r_{o}}\right)\right],\label{tor-grav-sf}
\end{equation}
where%
\begin{equation}
K_{i}=\frac{1}{2}\left|\frac{df}{dr}\right|_{r=r_i}.
\end{equation}
Explicitly, we have%
\begin{eqnarray}
  K_{e}=\frac{(r_{c}-r_{e})(r_{e}-r_{o})}{6r_{e}}\Lambda, \\
  K_{c}= \frac{(r_{c}-r_{e})(r_{c}-r_{o})}{6r_{c}}\Lambda,\\
  K_{o}= \frac{(r_{o}-r_{e})(r_{c}-r_{o})}{6r_{o}}\Lambda.
\end{eqnarray}
So under the tortoise coordinate transformation (\ref{tortoise }),
the
radial equation (\ref{radius equ. about r}) can be rewritten as%
\begin{equation}
\left[-\frac{d^2}{dx^2}+4M^2V(r)\right]\Psi_{\omega l n}
(x)=4M^2\omega^2\Psi_{\omega l n} (x),\label{radius-equation}
\end{equation}
which likes the form of Schr\"{o}dinger equation in quantum
mechanics. Notice that there are two various coordinates
--- $r$ and $x$ in this equation. So people also call it
Schr\"{o}dinger wavelike equation. The incoming or outgoing
particle flow between event horizon $r_{e}$ and cosmological
horizon $r_{c}$ is reflected and transmitted by the potential
$V(r)$. Substituting quantum parameters $\Omega_{n}$
(\ref{quan-Omega}) into Eq. (\ref{potential of r}), the quantum
potentials are obtained as follows
\begin{equation}\label{quan-potential}
    V_{n}(r)=f(r)\left[\frac{1}{r}\frac{df(r)}{dr}+\frac{l(l+1)}{r^2}+\frac{n^2\pi^2}{y_{1}^2}+\frac{3}{4}\Lambda\right].
\end{equation}
It is highly localizing near $r\sim (r_{e}+r_{c})/2\simeq 3$,
falling off exponentially in $x$ at both $r=r_{e}$ and $r=r_{c}$.
Comparing with the similar case of usual 4D \cite{ref:Brevik} two
additional monomials, $\frac{n^2\pi^2}{y_{1}^2}$ and
$\frac{3}{4}\Lambda$, have appeared in the potential. The form of
the potential for n=1, 2, 3 are illustrated in Fig.
\ref{fig:quan-potential-r}.

\subsection{The numerical solution}
Near the horizons $r_{e}$ and $r_{c}$,
$x\longrightarrow\pm\infty$. According to Eq. (\ref{re-f
function}) and Eq. (\ref{quan-potential}), we can get
\begin{equation}\label{Boundary-potential}
    V(r_{e})=V(r_{c})=0.
\end{equation}
So Eq. (\ref{radius-equation}) reduces to
\begin{equation}\label{Boundary-Equ}
    \left[\frac{d^2}{dx^2}+4M^2\omega^2\right]\Psi_{\omega l n} (x)=0.
\end{equation}
Its solutions are $e^{\pm i2M\omega x}$. In this paper, we only
take into account real field and choose the solution
\cite{ref:Brevik}
\begin{equation}\label{Boundary-conditions}
   \Psi_{\omega l n} =\cos(2M\omega x)
\end{equation}
as boundary condition near the two horizons (\ref{re-rc}).

\begin{figure}
 \centering
  \includegraphics[width=3.5in]{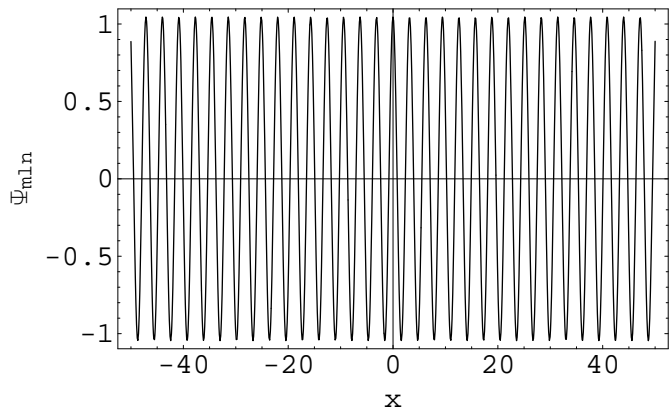}
  \caption{Variation of the field amplitude versus $x$ with M=1, l=1, $\Lambda=0.11$, $y_{1}=10^{3/2}$ and n=1. The solution is close to a harmonic wave.}\label{fig:0.11-x}
\end{figure}

\begin{figure}
 \centering
  \includegraphics[width=3.5in]{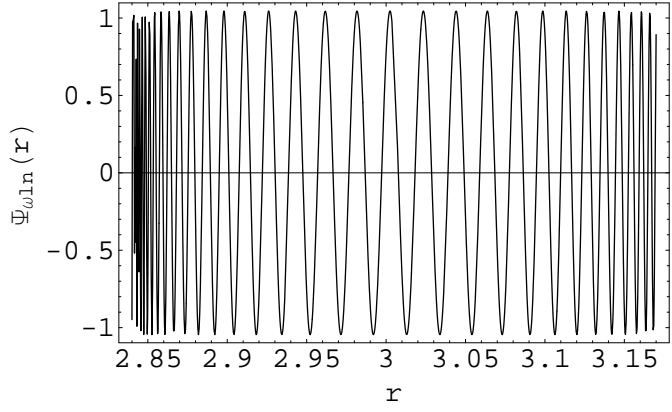}
  \caption{Variation of the field amplitude versus $r$ with M=1, l=1, $\Lambda=0.11$, $y_{1}=10^{3/2}$ and n=1. The waves pile up near the horizons.}\label{fig:0.11-r}
\end{figure}

In real scalar field case, there are two methods to solve
Schr\"{o}dinger wavelike equation (\ref{radius-equation}). One is
tangent approximation \cite{ref:Brevik} and the other is
polynomial approximation \cite{ref:Tian}. With any assigned
cosmological constant $\Lambda$, we can always find an appropriate
approximate method from those only by adjusting the parameters.
Here the former one is adopted to analyze this model. Hence, we
employe $\Lambda M^2 = 0.11$ and use the useful tangent
approximation \cite{ref:Brevik}
\begin{equation}
  \tilde{x}(r)=15\tan[b(r-d)+5],\label{tangent-appr}
\end{equation}
in which $b=2.7/(r_{c}-r_{e})$ and $d=(r_{c}+r_{e})/2$. Because
the approximation (\ref{tangent-appr}) does not allow $|x|$ to
become very large, we shorten the interval of $x$ to [-100,100].
So boundary condition (\ref{Boundary-conditions}) is rewritten as
\begin{equation} \label{100-boundary-condition}
    \Psi_{\omega l n} (-100)=\Psi_{\omega l n}
    (100)=\cos(200M\omega).
\end{equation}
Considering boundary condition (\ref{100-boundary-condition}) and
tangent approximation (\ref{tangent-appr}), we can solve Eq.
(\ref{radius-equation}) numerically as a boundary value problem by
Mathematica software. The variation amplitude of waves
$\Psi_{\omega l}$ versus tortoise coordinate $x$ is illustrated in
Fig. \ref{fig:0.11-x}. Considering actual circumstance, we use
tortoise transformation (\ref{tor-grav-sf}) and also plot the
amplitude versus $r$ in Fig. \ref{fig:0.11-r}, where we only give
the first quantum state (n=1). The others can be treated by the
same way.

\section{the reflection and transmission}
We assume that the particle flux with energy $E$ bursts towards a
square well along the positive direction of $x$ axis, where the
potential is
\begin{equation}\label{square-well}
    \hat{V}(x)=\left\{
\begin{array}{c}
V_{0}, \ \ \ \ x_{1}<x<x_{2},\\
\ \ \ 0, \ \ \ x<x_{1}\  \text{or}\  x>x_{2}.\\
\end{array}
\right.
\end{equation}
From the view of quantum mechanics, considering the wave behavior
of the particles, this process is similar to scattering on the
surface of propagation medium with thickness of $|x_{2}-x_{1}|$.
Parts of them are transmitted and parts of them are reflected
back. According to statistical interpretation of wave function,
whether the energy $E > V_{0}$ or not, there is definite
probabilities to transmit or reflect by the potential. The
reflection and transmission coefficients denote the magnitude of
those probabilities.

\begin{figure}
\centering
\includegraphics[width=3.5in]{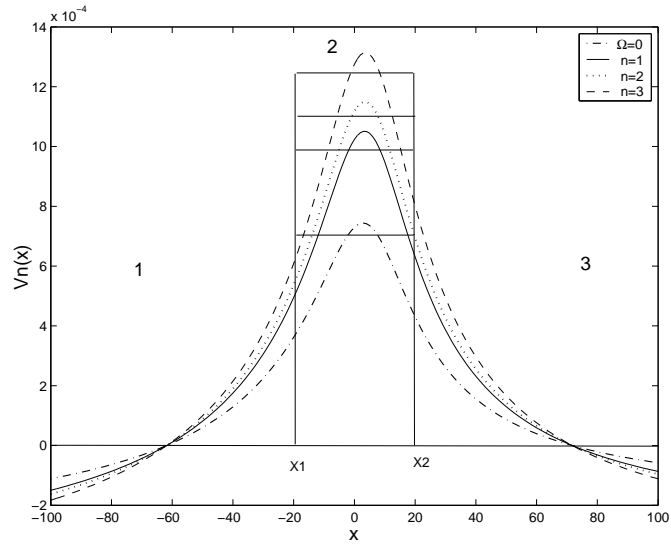}
\caption{Replacement of real 5D SdS potential barriers around
Nariai black hole by square barriers with n=1 (solid), n=2
(dotted), n=3 (dashed) and $\Omega=0$ (usual 4D case with dash-dot
line). We use M=1, l=1, $\Lambda=0.11$ and
$y_{1}=10^{3/2}$.}\label{fig:replacement}
\end{figure}

As mentioned above, it is necessary to replace the continuously
varying potential barrier with a discontinuous barrier of constant
height in analytical work. Therefore, the usual reflection and
transmission coefficients can be obtained. With the method of
\cite{ref:Stratton} and \cite{ref:Brevik}, We suppose a scalar
wave propagates from $-\infty$ to $+\infty$, which is illustrated
in Fig. \ref{fig:replacement}. The same denotation is cited here,
namely associating $''1''$ with the incoming wave in the region
$-\infty < x < x_{1}$, $''2''$ with the potential plateau $x_{1} <
x < x_{2}$, and $''3''$ with the outgoing wave in the region
$x_{2} < x <+\infty$. Hence, potential $V(x)$ in Eq.
(\ref{radius-equation}) reduces to
\begin{equation}\label{sp}
    V(x)=\left\{
\begin{array}{c}
\hat{V}_{1},\ -\infty<x<x_{1},\\
\hat{V}_{2},\ \ \ x_{1}<x<x_{2},\\
\ \ \hat{V}_{3},\ \ \ x_{2}<x<+\infty.\\
\end{array}
\right.
\end{equation}

\begin{figure}
\centering
\includegraphics[width=3.5in]{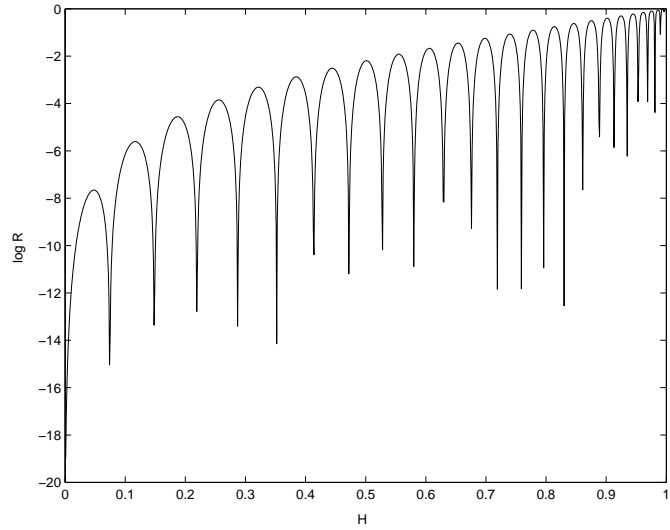}
\caption{$\log R$ versus height H (or $\hat{V_{2}}$) with the
width $d = 40$, $M=1$, $l=1$, $\Lambda=0.11$ and
$y_{1}=10^{3/2}$}\label{logR_H}
\end{figure}

According to square barrier (\ref{sp}), the solutions to
Eq.(\ref{radius-equation}) are
\begin{equation}\label{solu-p}
    \Psi_{\omega l n}=\left\{
\begin{array}{c}
a_{1}e^{ik_{1} x}+b_{1} e^{-ik_{1} x},\ -\infty<x<x_{1},\\
a_{2} e^{i k_{2} x}+b_{2} e^{-ik_{2} x},\ \ \ x_{1}<x<x_{2},\\
a_{3} e^{ik_{3}x},\ \ \ \ \ \ \ \ \ \ \ \ \ \ \ \ x_{2}<x<+\infty,\\
\end{array}
\right.
\end{equation}
where $k_{i}=\sqrt{4M^2(\omega^2-\hat{V}_{i})}$ ($i=1,2,3$) are
the wave numbers; $a_{i}$ and $b_{i}$ are the undetermined
coefficients to the solutions . Then we define reflection
coefficients for the plane interfaces dividing two media
\begin{equation}\label{Rij}
    R_{ij} = \left(\frac{1-Z_{ij}}{1+Z_{ij}}\right)^2,
\end{equation}
where $Z_{ij} = \frac{k_{j}}{k_{i}}$ are the real impedance ratios
between medium $i$ and $j$. The width of barrier is
$d=x_{2}-x_{1}$ and the height of square barrier is $H =
\hat{V}_{2}$. So in this model reflection coefficients $R$ and
transmission coefficients $T$ are given as
\begin{eqnarray}
R&=&\left|\frac{b_{1}}{a_{1}}\right|^2=\frac{R_{12}+R_{23}+2\sqrt{R_{12}R_{23}}\cos(2k_{2}d)}{1+R_{12}R_{23}+2\sqrt{R_{12}R_{23}}\cos(2k_{2}d)} ,\label{R}\\
T&=&\left|\frac{a_{3}}{a_{1}}\right|^2=\frac{1}{(1+Z_{12})^2(1+Z_{23})^2}
\frac{16}{1+R_{12}R_{23}+2\sqrt{R_{12}R_{23}}\cos(2k_{2}d)}.\label{T}
\end{eqnarray}
Because the same width are adopted here, we only give the
functional image of $\log R$ versus hight H (or $\hat{V_{2}}$) in
Fig. \ref{logR_H}. There is no surprise that it take oscillating
like forms. One can read this feature directly from Eq. (\ref{R}),
which contains cosine functions. Then we use tangent approximation
(\ref{tangent-appr}) and get the replacements of the 5D continuous
potentials by square barriers in Fig. \ref{fig:replacement}. The
different reasonable $\hat{V_{2}}$ are read off as the height of
those square barriers. Meanwhile, we choose the incoming wave
number to be $k_{1} = 2$ ($\hat{V_{1}} = 0$) and $k_{3} = 2$
($\hat{V_{3}} = 0$). Substituting those parameters into Eqs.
(\ref{R}) (\ref{T}), we can obtain reflection and transmission
coefficients. Comparing those coefficients with usual 4D SdS case
\cite{ref:Brevik}, one can see the difference clearly in Table
\ref{table}. Viewing from Fig. \ref{logR_H}, we can get a
relationship of the four heights $H_{\Omega = 0} < H_{n = 1} <
H_{n = 2} < H_{n = 3} < H_0$ (\text{the horizontal ordinate of the
first extreme point}). Hence, we can say that the four modes
($\Omega = 0, \ n=1, \ n=2, \ n=3$) are in the same monotone
increasing space. Obviously, the reflection coefficients $R$ (or
$T$, notice $R + T = 1$) of 5D SdS black string are bigger (or
smaller) than 4D case. Else, $R|_{n=1} < R|_{n=2} < R|_{n=3}$.

\begin{table}[!h]
\tabcolsep 0pt \caption{The reflection and transmission
coefficients} \vspace*{-12pt}
\begin{center}
\def\temptablewidth{0.5\textwidth}
{\rule{\temptablewidth}{1pt}}
\begin{tabular*}{\temptablewidth}{@{\extracolsep{\fill}}ccccccc}
      mode &$x_{1}$&$x_{2}$&d&$v_{2}$ (or H)&R \\\hline
     4D SdS&-20&20&40&$7.3\times 10^{-4}$&$1.3\times 10^{-7}$\\
     n=1&-20&20&40&$10\times 10^{-4}$&$2.5\times10^{-7}$\\
     n=2&-20&20&40&$11\times 10^{-4}$&$3.0\times10^{-7}$\\
     n=3&-20&20&40&$12.5\times 10^{-4}$&$3.9\times10^{-7}$
       \end{tabular*}\label{table}
       {\rule{\temptablewidth}{1pt}}
       \end{center}
       \end{table}

\section{conclusion}
In this paper we have solved the real scalar field $\Phi$ and
obtained reflection and transmission coefficients ($R$, $T$)
around the Nariai black hole in the 5D SdS black string space. We
summarize what have been achieved.

1. The 5D solution presented by Mashhoon, Wesson  and Liu
\cite{ref:Liu} \cite{ref:Mashhoon} \cite{ref:Wesson} is exact in
higher dimensional gravity theory. It satisfies the 5D Ricci-flat
field equation $R_{AB}=0$.  In this paper, two branes are embedded
into the bulk. One brane is at $y =0$ where the standard matter
lives. The other brane is at $y = y_ {1} $, where $y_{1}$ is the
thickness of bulk. It is the basal topological structure of this
black string space. The usual 4D effective cosmological constant
$\Lambda$ is considered to be induced from the 5D Ricci-flat
space. One should notice that $\Lambda$ is considered as a free
parameter. The distance between black hole horizon and
cosmological horizon is shorten with bigger $\Lambda$. In this
metric it has nothing to do with the value referred from current
cosmological observation. So, if the value of $\Lambda$ increases,
the Nariai black hole is inevitably arisen in its last fate.

2. For the well known Nariai case, we have solved the scalar field
around it in the 5D SdS black string space.  The fifth dimensional
component $L(y) $ is presented. Taking into account the classical
field theory, one know that standard model fields (such as
fermions, gauge bosons, Higgs fields) are confined on a $(3 + 1)$
dimensional hypersurface (3-brane) without accessing along the
transverse dimensions. In order to stabilize two branes, the
scalar field is led in. According to $''\text{standing wave}''$
condition, the fifth dimensional equation can be solved. The
scalar field gets its maximum on the brane $y = 0$ and get an
extremeness value on the other brane $y = y_{1}$. So the spectrum
of parameter $\Omega$ is broken into two parties, one is quantum
$\Omega_{n}$ and the other is continuous one. The quantum spectrum
is illustrated in Fig. \ref{quan-states}. It is clear that the
extreme Nariai black hole decays more acutely than usual case
\cite{ref:Liu00}. Furthermore, the quantum phenomenon emerges
distinctly in the waves. One can see those according to the
effective potential (\ref{quan-potential}), Fig. \ref{fig:0.11-x}
and Fig. \ref{fig:0.11-r}.

3. Because of the singular $f(r) = 0$ in the metric, potential
(\ref{potential of r}) vanishes both near black hole horizon and
cosmological horizon. Then the Schr\"{o}dinger wavelike equation
(\ref{radius-equation}) reduces to a solvable one
(\ref{Boundary-Equ}). Obviously, according to the real scalar
field, we get the boundary conditions (\ref{Boundary-conditions}).
Eq.(\ref{radius-equation}) describes one dimensional transmission
of waves scattering by potential barrier. In order to solve this
equation, we adopt a useful tangent approximation
\cite{ref:Brevik} to unite radial coordinate $r$ and tortoise
coordinate $x$. So Eq. (\ref{radius-equation}), effective
potential (\ref{quan-potential}) and boundary condition
(\ref{Boundary-conditions}) constitute a full boundary value
problem. Because of the complicated potential
(\ref{quan-potential}) and fitting function (\ref{tangent-appr}),
we only give numerical solution. By used the replacement, the
continuous potential is switched into a square barrier. With the
classical method of the square barrier, the reflection and
transmission coefficients ($R, \ T$) are obtained naturally. The
result is presented briefly in Table \ref{table}.

\acknowledgments

We are grateful to Li Chen for useful help and also appreciate
Feng Luo's good advice. This work was supported by NSF (10573003)
and NBRP(2003CB716300) of P. R. China.

\end{document}